\documentstyle[12pt]{article}
\begin{document}
\setlength{\baselineskip}{22pt}

\vspace{6mm}

\begin{center}
{\Large \bf Comment on `` Fractal formation and ordering in Random 
Sequential Adsorption"} \end{center}

\vspace{5mm}

\begin{center}
{\bf M. K. Hassan}
\end{center}
\begin{center}
{\bf Department of Physics, Brunel University, Uxbridge, Middlesex UB8
  3PH, United Kingdom}
\end{center}
\begin{center}
E-mail:Kamrul.Hassan@brunel.ac.uk
\end{center}
           
\vspace{5mm}
             
In a recent letter [1], Brilliantov {\it et al.} studied the random 
sequential adsorption (RSA) of a mixture of particles with continuous 
distribution of sizes. In particular, they choose a power law form for the 
parking distribution, $p(x) \sim \alpha x^{\alpha -1}$, and find the 
resulting pattern that arises in the long time can be described 
geometrically 
using the fractal concept. They also claim that the pattern becomes more 
and more ordered as $\alpha$ increases and reaches to the Apollonian 
packing in 
the limit $\alpha \rightarrow \infty$. This finding does not seem to be 
correct at least in one dimension. Here we attempt to find an exact 
enumeration of the fractal dimension $D_f$ analytically in one dimension for 
different values of $\alpha$ and give the physical 
interpretation of the role played by the exponent $\alpha$.

Defining $c(x,t)$ as the concentration of empty spaces of size $x$ at 
time $t$, the RSA process in one dimension can be described by the linear 
integro-differential equation [2]
\begin{eqnarray}
{{\partial c(x,t)}\over{\partial t}} 
& = & -c(x,t)\int_0^xp(z)dz\int_0^{x-z}F(y,x-y-z|z)dy+ \\ \nonumber 
& & 2\int_x^\infty c(y,t)dy\int_0^{y-x}p(z)F(x,y-x-z|z)dz
\end{eqnarray}
where, $F(x,y|z)$ determine the rate at which particles attempt to 
deposit with the probability $p(z)$. For a power-law form of 
$p(z)$ and for uniform deposition rate $F(x,y|z)=1$, the rate 
equation for the $qth$ moment [3] is 
\begin{equation}
{{dM_q(t)}\over{dt}} =\Big [ 
{{2\Gamma(\alpha+2)\Gamma(q+1)}\over{\Gamma(q+\alpha+2)}}-1\Big 
]M_{q+\alpha+1}(t),
\end{equation}
where, $M_q(t)=\int_0^\infty c(x,t)x^qdx$.
We now consider a system in which an arbitrary number of particles, $n$, 
are created  and $(n-m)$ of them are removed at each time step and the 
process is repeated {\it ad infinitum}, where,  $2\leq m<n$. The rate 
equation for $c(x,t)$ is [2] 
\begin{eqnarray}
{{\partial c(x,t)}\over{\partial t}} & = & 
-c(x,t)\int\prod_{j=1}^{n-m}p
(x_j)F(x_{n-m+1},..,x_n|x_1,..,x_{n-m})\delta(x-\sum_{i=1}^nx_i)\prod_{i=1}^n
dx_i +m   \nonumber \\ &  & \int
c(y,t)F(x,x_{n-m+1},..,x_{n-1}|x_1,..,x_{n-m})\delta(y-x-
\sum_{i=1}^{n-1}x_i)\prod_{i=1}^{n-1}dx_idy\prod_{j=1}^{n-m}p(x_j).
\nonumber \\
\end{eqnarray}
Choosing $p(x_i)\sim (\beta+1)x_i^\beta$ and 
$F((x_{n-m+1},..,x_n|x_1,..,x_{n-m})=x_{n-m+1}^\beta...x_n^\beta$ we 
obtain 
\begin{equation}
{{dM_q(t)}\over{dt}}=\Big ({{m\Gamma(q+\beta+1)}\over
{\Gamma(q+n(\beta+1))}}-{{\Gamma(\beta+1)}\over{\Gamma(n(\beta+1))}}\Big )
M_{q+n(\beta+1)-1}(t)
\end{equation}
Note that if we set $\beta=0, m=2$ and $n=\alpha+2$ 
we get exactly the same equation as equation (2) this implies that the 
pattern 
created by this two systems are identical. It also  implies that at each 
time step a 
line is broken into $(\alpha +2)$ pieces and $\alpha$ of them are removed. 
Equivalently, we can consider the deposition of 
one particle at each event whose length is the same as  the length of 
$\alpha$ number of discarded particles. Thus, as $\alpha$ increases the 
length of the deposited particle increases on the average. In order to 
find the fractal dimension, we  now define a line segment 
$\delta \sim {{M_1(t)}\over{M_0(t)}}$ and find the number of segments 
required to cover the pattern created scales as $N(\delta) \sim  
\delta^{-D_f}$ in the limit $\delta \rightarrow 0$. The fractal 
dimension,  $D_f$ is the real and positive root  of the polynomial 
equation in $q$ obtained by setting the parenthesis of 
equation (2) equal to zero [2] for which the moment becomes time 
independent. A detailed numerical survey 
confirms that the fractal dimension decreases as $\alpha$ increases (see 
the Table I).
\vspace{3mm}

\begin{center}
\begin{tabular}{|l|l|l|}\hline
$\alpha$ & $\beta=0$ & $\beta=1$ \\ \cline{1-3}
$1$ & $0.5616$ & $0.5956$ \\  \hline
$2$ & $0.4348$ & $0.5183$   \\  \hline
$3$ & $0.3723$ & $0.466542$   \\ \hline
$4$ & $0.33405$ & $0.429121$   \\  \hline
$5$ & $0.30784$ & $0.400614$  \\ \hline
$6$ & $0.288505$ & $0.37805$  \\ \hline
$7$ & $0.27351$  & $0.35966$  \\ \hline
\hline
\end{tabular}
\end{center}
Table I: The fractal Dimension $D_f$ for $\beta =0,1$ for different
values of $\alpha$.

 \vspace{3mm}

\noindent 
However, if we choose $p(z) \sim (\alpha+1)z^\alpha$ and 
$F(x,y|z) =(xy)^\alpha$ in equation (1) then we can  get the rate equation 
for the moment from equation (4) by setting $n=3, m=2$ and $\beta=\alpha$. 
For this particular choice we find that the fractal dimension 
increases as $\alpha$ increases and reaches to ${{\ln 2}\over{\ln 3}}$ as 
$\alpha \rightarrow \infty$ (see table II), which is the 
fractal dimension for the classic Cantor set (non-random fractal). In 
this case the $\alpha$ value determines the degree of tendency to place the 
particles in the centre and in the limit $\alpha \rightarrow 
\infty$ particles are always placed exactly in the centre of the empty 
space. 

\vspace{3mm}

\begin{center}
\begin{tabular}{|l|l|} \hline
 $\beta$ & $ D_f$ \\
\hline
$-{{1}\over{2}}$ & ${{1}\over{2}}$ \\
0 & 0.5616288 \\
${{1}\over{2}}$ & 0.5841 \\
1 & 0.5956 \\
2 & 0.6073 \\
$\infty$ & 0.6309 \\
\hline
\end{tabular}
\end{center}
Table 2: The fractal dimension $D_f$ of stochastic fractals for
$m=2$ and $n=3$ for increasing $\beta=\alpha$ values when deposition is 
position dependent. 

\vspace{3mm}

In other 
words, the $\alpha$ value determines the degree of 
order. It shows that the  power-law size distribution  alone does not 
produce any ordered packing, instead, one has to chose the deposition 
rate,  $F(x,y|z)$, to be position dependent following the similar power-law 
form as for $p(z)$ for which in the limit $\alpha \rightarrow \infty$ 
the pattern becomes ordered. We conclude with the remark that the fractal 
dimension increases with the degree of increasing order and 
reaches  it's maximum value in the perfectly ordered pattern as it 
does in the classic Cantor set or in the Sierpinski gasket.

\vspace{3mm}

\noindent
The author would like to thank the CVCP for an ORS award.

\vspace{5mm}


\vspace{3mm}

\noindent
[1] N. V. Brilliantov, Yu. A Andrienko, P. L Krapivsky and J. 
Kurths,
 
\noindent
Phys. Rev. Lett. {\bf 76} 4058 (1996).

\noindent
[2] M.  K. Hassan and G. J. Rodgers, Phys. Lett. A {\bf 208} 95 (1995).

\noindent
[3] P. L. Krapivsky, J. Stat. Phys. {\bf 69} 135 (1992).

\end{document}